\documentclass[aps,prd,reprint,preprintnumbers,floats,epsfig,nofootinbib,amssymb,
nofootinbib]{revtex4}

\usepackage{slashed}
\usepackage{graphicx}
\usepackage[dvipsnames]{xcolor}

\usepackage{epstopdf}
\usepackage{dcolumn}
\usepackage{bm}
\usepackage{color}
\usepackage{ulem}
\usepackage{enumitem}
\usepackage{amsmath}
\usepackage[compat=1.1.0]{tikz-feynman}
\usepackage{multirow}
\usepackage{graphicx}
\usepackage{subcaption}
\usepackage{url}
\usepackage{mathrsfs}
\usepackage{indentfirst}
\usepackage{float}
\usepackage[colorlinks,
linkcolor=red,
anchorcolor=blue,
citecolor=blue
]{hyperref}
\usepackage{cancel}
\usepackage{amssymb}
\usepackage{textcomp}
\usepackage{amsmath}
\usepackage{bm}
\usepackage[T1]{fontenc}
\usepackage{times}
\usepackage{epsfig}

\begin{document}

	\baselineskip=15pt
	
	\preprint{CTPU-PTC-25-38}

\title{
Collider Probes of Four-Lepton Final States in  Maximally Flavor-Violating $U(1)_{L_\mu - L_\tau}$ Model
}

\author{Jianing Qin${}^{1}$}

\author{Fei Huang${}^{1}$\footnote{sps\_huangf@ujn.edu.cn}}
\author{Honglei Li${}^{1}$\footnote{sps\_lihl@ujn.edu.cn}}
\author{Zhi-Long Han${}^{1}$\footnote{sps\_hanzl@ujn.edu.cn}}
\author{ Jin Sun$^{2}$\footnote{sunjin0810@ibs.re.kr}}

\affiliation{${}^{1}$School of Physics and Technology, University of Jinan, Jinan, Shandong 250022, China}

\affiliation{${}^{2}$Particle Theory and Cosmology Group, Center for Theoretical Physics of the Universe, Institute for Basic Science (IBS), Daejeon 34126, Korea}

\date{\today}

\begin{abstract}

We investigate the collider signatures of the maximally flavor-violating  $U(1)_{L_\mu-L_\tau}$
model, where a new gauge boson $Z^\prime$
 and scalar triplets induce lepton flavor-changing interactions in the $\mu$-$\tau$ sector.
Focusing on four-lepton final states at multi-TeV lepton colliders, we conduct a detailed analysis of cross sections, asymmetries, and polarization effects.
We show that the signal cross section is highly sensitive to $m_{Z^\prime}$ and the effective parameters $\tilde{g}/m_{Z^\prime}$, 
while remaining largely insensitive to the triplet Yukawa couplings within the phenomenologically allowed region.
The forward-backward asymmetry exhibits a characteristic monotonic dependence on $m_{Z^\prime}$, and beam polarization can significantly suppress Standard Model backgrounds while enhancing new physics contributions.
We find that over the phenomenologically allowed parameter space, the predicted observables remain highly sensitive to the underlying model parameters. These results demonstrate that multi-lepton final states are powerful probes of the $U(1)_{L\mu - L_\tau}$ framework and offer valuable guidance for future searches at muon and electron-positron colliders.

\end{abstract}

\maketitle

\section{Introduction}

The Standard Model (SM) of particle physics successfully describes all known elementary particles and their interactions~\cite{Rosner:2001zy}. 
However, the observation of neutrino oscillations implies that neutrinos possess nonzero masses and that neutral lepton flavor violation (LFV) occurs, thereby strongly indicating the potential for charged lepton flavor violation. 
Within the SM, charged LFV processes arise only at the loop level and are highly suppressed due to the minuscule neutrino masses compared to the electroweak scale, with branching ratios such as $Br(\mu \to e\gamma)\sim 10^{-55}$~\cite{Calibbi:2017uvl,Jahedi:2025hnu}. Consequently, the predicted rates of LFV processes within the SM lie far below the reach of current and foreseeable experimental sensitivity.
Therefore, any observable LFV signal would constitute compelling evidence for physics beyond the SM.

Numerous extensions of the SM predict observable charged LFV, including supersymmetric models~\cite{Hirsch:2008gh,Calibbi:2017uvl,Huang:2024ozb}, left-right symmetric frameworks~\cite{Das:2012ii,Barry:2013xxa,Deppisch:2014zta,Bambhaniya:2015ipg,Bonilla:2016fqd}, various seesaw mechanisms~\cite{Chowdhury:2013jta,He:2014efa,BhupalDev:2018tox,Crivellin:2022cve}, and leptoquark scenarios~\cite{Davidson:1993qk,Dorsner:2016wpm,Varzielas:2023qlb,De:2024foq}, Goldstone scenarios~\cite{Cheng:2020rla,Sun:2021jpw} and $Z'$ models~\cite{Foot:1994vd,Cen:2021ryk,Cheng:2021okr,Huang:2024iip,Liu:2024gui}, etc., each providing
distinct sources and characteristic signatures of flavor violation.
Among these, the anomaly-free $U(1)_{L_\mu - L_\tau}$ gauge symmetry is particularly compelling, as it naturally explains the flavor structure of the lepton sector\textcolor{red}{~\cite{He:1990pn,He:1991qd,Rodejohann:2005ru,Heeck:2011wj,Asai:2018ocx,Majumdar:2020xws,Heeck:2022znj}}. 
In this work, we investigate a maximally flavor-violating realization of the $U(1)_{L_\mu - L_\tau}$ model, which permits sizable LFV couplings and leads to distinctive multilepton signatures at colliders.
This framework is realized by introducing off-diagonal gauge interactions and new scalar fields that break the flavor symmetry in a controlled manner~\cite{He:1994aq}.
 As a result, the associated $Z'$ boson  and 
 extended Higgs sector can mediate LFV processes at tree level, significantly enhancing the prospects for observing such signals.
Furthermore, the model remains
 consistent with existing constraints from precision electroweak measurements, neutrino oscillation data, and rare decay searches, provided that the $Z'$
 mass and coupling strengths are appropriately chosen.
 Notably, sizable LFV couplings are allowed without violating current experimental limits, particularly in the $\mu \text{-} \tau$ sector, which is significantly less constrained than the 
$e \text{-} \mu$ and $e \text{-} \tau$ channels~\cite{ATLAS:2020tre}.

Lepton colliders offer a clean experimental environment with well-defined initial states, making them ideal for precision electroweak measurements and the search for rare or forbidden phenomena such as lepton flavor violation (LFV)~\cite{Porod:2002zy,Li:2018cod,Dev:2017ftk,Qin:2017aju,Fridell:2023gjx,Sun:2023ylp,Altmannshofer:2023tsa,Li:2025prq}.
In particular, future high-luminosity electron-positron colliders—such as the  CLIC~\cite{Aicheler:2018arh}, ILC~\cite{Barklow:2015tja}, CEPC~\cite{CEPCStudyGroup:2023quu,Ai:2024nmn}, and FCC-ee~\cite{FCC:2018evy}—will provide excellent sensitivity to multilepton final states, owing to their low background levels and high lepton detection efficiencies.
These facilities enable detailed reconstruction of kinematic distributions and invariant mass spectra, thereby facilitating resonance searches for new particles such as the $Z'$ boson.
Within the framework of the $U(1)_{L_\mu - L_\tau}$
 model with maximal LFV, the observation of final states involving mixed lepton flavors~\cite{Huang:2025osf} would constitute a smoking-gun signature of new physics.
 Moreover, advanced flavor-tagging techniques and high energy resolution enable the disentanglement of various decay modes and a detailed study of the underlying LFV couplings.
In addition to electron-positron colliders, muon colliders have recently attracted increasing attention as promising next-generation facilities.
 With the potential to achieve multi-TeV center-of-mass energies and 
 deliver high luminosity, muon colliders offer powerful probes of both flavor-conserving and flavor-violating new physics~\cite{MuonCollider:2022xlm}. 
 Their unique initial state allows for the direct production of particles that couple to muons, such as the $Z'$ boson in the $U(1)_{L_\mu - L_\tau}$ model~\cite{Sun:2023ylp}. 
 Correspondingly,  the large muon coupling enhances sensitivity to interactions involving second- and third-generation leptons, making muon colliders particularly well-suited for probing LFV in the $\mu \text{-} \tau$ sector.

In this paper, we focus on four-lepton final states as a clean and sensitive probe of lepton flavor violation. The structure of the paper is as follows. In Section \ref{section2}, we present the construction of the maximally flavor-violating $U(1)_{L_\mu - L_\tau}$ model, where the $Z'$ interactions are augmented by triplet scalar fields. 
In Section \ref{section3},
we analyze the resulting collider signatures, with a focus on four-lepton final states at both muon and electron-positron colliders.
 Finally, we summarize our findings and offer concluding remarks in Section \ref{section4}.

\section{The $U(1)_{L_\mu-L_\tau}$  model for maximal $\mu-\tau$ coupling}\label{section2}

The $U(1)_{L_\mu-L_\tau}$ gauge symmetry corresponds to the difference between the muon and tau lepton numbers.
In the minimal setup, the left-handed doublets, $L_{L\;i}: (1, 2, -1/2)$, and the right-handed singlets, $e_{R\;i}: (1,1,-1)$, transform under the gauged $U(1)_{L_\mu-L_\tau}$ symmetry with charges $0$, $+1$, and $-1$
assigned to the first, second, and third generations, respectively.
Here, the numbers in parentheses denote the quantum numbers under the $SU(3)_C\times SU(2)_L\times U(1)_Y$ gauge group.
The $Z^\prime$ gauge boson in this model couples exclusively to leptons in the weak interaction basis~\cite{He:1990pn, He:1991qd}
\begin{eqnarray}\label{conserving}
	{\cal L}_{Z'}=- \tilde g (\bar \mu \gamma^\mu \mu - \bar \tau  \gamma^\mu \tau + \bar \nu_\mu \gamma^\mu L \nu_\mu - \bar \nu_\tau \gamma^\mu L \nu_\tau) Z^\prime_\mu \;, \label{zprime-current}
\end{eqnarray}
where $\tilde g$ is the gauge coupling associated with the $U(1)_{L\mu - L_\tau}$ symmetry.
And $L/R$ denotes the chiral projection $P_{L/R}=(1\mp \gamma_5)/2$.
The vector current interaction provides a positive contribution to the muon anomalous magnetic moment $(g-2)_\mu$.
The $U(1)_{L_\mu - L_\tau}$ symmetry is spontaneously broken by introducing a singlet scalar field $S$ with charge $+1$ and a vacuum expectation value (VEV) of $v_S/\sqrt{2}$, through which the $Z^\prime$ boson acquires a mass given by $m_{Z^\prime} = \tilde g v_S$.

The interaction described above is purely flavor-conserving but can be extended to incorporate flavor-changing effects.
We briefly outline the construction of such a model below.
This extension requires the introduction of three Higgs doublets, $H_{1,2,3}: (1,2, 1/2)$, with VEVs $\langle H_i \rangle = v_i/\sqrt{2}$, and $U(1)_{L_\mu - L_\tau}$ charges $0$, $+2$, and $-2$, respectively.
An unbroken exchange symmetry is imposed under which $Z^\prime \to - Z^\prime$, $H_1 \leftrightarrow H_1$, and $H_2 \leftrightarrow H_3$, with $v_2 = v_3 \equiv v$.
Under these conditions, the $Z^\prime$ interactions and lepton Yukawa couplings are given by               \begin{eqnarray}\label{yukawa}
	{\cal L}_{H}= &&- \tilde g (\bar l_2 \gamma^\mu L l_2- \bar l_3  \gamma^\mu L l_3 + \bar e_2 \gamma^\mu R e_2 - \bar e_3 \gamma^\mu R e_3) Z^\prime_\mu\nonumber\\
	&&- [Y^l_{11} \bar l_1 R e_1 + Y^l_{22} (\bar l_2 R e_2 +\bar l_3 R e_3 ) ] H_1 
	-Y^l_{23} (\bar l_2 R e_3 H_2 +\bar l_3 R e_2 H_3 ) +  \text{H.C.}
\end{eqnarray}
After electroweak symmetry breaking, the scalar fields acquire non-zero VEVs, which generate the charged lepton masses via the Yukawa interactions shown in the second line of Eq.~(\ref{yukawa}).
However, since the resulting mass matrix is not diagonal, it must be diagonalized through the following transformation:
\begin{eqnarray}\label{eigen}
	\left (
	\begin{array}{c}
		\mu\\
		\tau
	\end{array}
	\right )
	= \frac{1}{\sqrt{2}}
	\left (
	\begin{array}{rr}
		1&\;-1\\
		1&\;1
	\end{array}
	\right )
	\left (\begin{array}{c}
		e_2\\
		e_3
	\end{array}
	\right ),\;
	\end{eqnarray}
A similar transformation is also applied to the neutrino sector.
After rotating to the charged lepton mass eigenstate basis using the above transformation, the desired flavor-changing $Z^\prime$ interactions take the following form
	\begin{eqnarray}
		{\cal L}_{Z'}=	- \tilde g (\bar \mu \gamma^\mu \tau +  \bar \tau  \gamma^\mu \mu + \bar \nu_\mu \gamma^\mu L \nu_\tau  + \bar \nu_\tau \gamma^\mu L \nu_\mu) Z'_\mu \;. \label{zprime-changing}
	\end{eqnarray}
The aforementioned $Z^\prime$ interaction significantly enhances the branching ratio of the process $\tau \to \mu \bar{\nu}_\mu \nu_\tau$, an effect that can be alleviated by introducing three scalar triplets with hypercharge $Y=1$, $\Delta_{1,2,3}: (1,3,1)$, each acquiring VEVs $\langle \Delta_i \rangle = v_{\Delta i}/\sqrt{2}$ and carrying $U(1)_{L_\mu-L_\tau}$ charges of $0$, $-2$, and $+2$, respectively~\cite{Cheng:2021okr}.
The $\Delta$ fields form the core of the well-known type-II seesaw mechanism, which generates small neutrino masses~\cite{Lazarides:1980nt,Mohapatra:1980yp,Konetschny:1977bn,Cheng:1980qt,Magg:1980ut,Schechter:1980gr} with 
	\begin{eqnarray}
		\Delta = \left (\begin{array}{cc}   \Delta^+/\sqrt{2}&\;  \Delta^{++}\\  \Delta^0&\; - \Delta^+/\sqrt{2} \end{array} \right )\;,\;\;\;\Delta^0=\frac{v_\Delta+\delta+i\eta}{\sqrt{2}}\;. 
	\end{eqnarray}
Under the above exchange symmetry $\Delta_1 \leftrightarrow  \Delta_1,\; \Delta_2 \leftrightarrow  \Delta_3$ with $v_{\Delta 2} = v_{\Delta 3}$, we use the transformation given in Eq.~(\ref{eigen}) to express the Yukawa terms in terms of the component fields $\Delta^{0}$, $\Delta^{+}$, and $\Delta^{++}$ as
	\begin{eqnarray}
		&&
		{\cal L}_\Delta = - (\bar \nu_e^c, \bar \nu_\mu^c, \bar \nu_\tau^c ) M(\Delta^0)  L
		\left ( \begin{array}{c}
			\nu_e \\ \nu_\mu \\ \nu_\tau
		\end{array}
		\right )
		+ \sqrt{2} (\bar \nu_e^c, \bar \nu_\mu^c, \bar \nu_\tau^c ) M(\Delta^+) L
		\left ( \begin{array}{c}
			e \\ \mu \\ \tau
		\end{array} \right )
		+ (\bar e^c, \bar \mu^c, \bar \tau^c ) M(\Delta^{++}) L
		\left ( \begin{array}{c}
			e \\ \mu \\ \tau
		\end{array} \right )
		\;,\nonumber\\
		&&\mbox{with}\;\;\; M(\Delta)  = \left ( \begin{array}{ccc}
			Y^\nu_{11} \Delta_1&0&0\\
			\\
			0& (Y^\nu_{22}(\Delta_2 +\Delta_3) - 2 Y^\nu_{23}\Delta_1)/2 & Y^\nu_{22}(\Delta_2-\Delta_3)/2\\
			\\
			0&Y^\nu_{22}(\Delta_2-\Delta_3)/2&(Y^\nu_{22}(\Delta_2 +\Delta_3) + 2 Y^\nu_{23}\Delta_1)/2
		\end{array}
		\right ) \;.
	\end{eqnarray}
We find that the large number of model parameters complicates the analysis. For simplicity, we adopt the following assumptions.
\begin{itemize}
\item \textbf{$ m_{\Delta^{++,+,0}}= m_\Delta$}: The degeneracy of the triplet components guarantees that the triplet scalars possess identical masses, a requirement that demands $m_\Delta > 420$ GeV~\cite{Ashanujjaman:2021txz} for $v_\Delta \sim \mathcal{O}(1)$ GeV.
\item \textbf{$Y_{11,23}<<Y_{22}$}: 
This suppresses the effects of $\Delta_1$, resulting in triplet contributions dominated by $\Delta_{2}$ and $\Delta_{3}$.
\item \textbf{ $\Delta_2\sim \Delta_3$}: 
Off-diagonal interactions are forbidden by the exact exchange symmetry, which may be broken by quantum corrections at the two-loop level. Therefore, we consider the case where 
the diagonal coupling is approximately ten orders of magnitude larger than the off-diagonal coupling.
\end{itemize}
The above assumptions ensure that the triplet effects can be effectively described solely by the $\Delta_2$ parameters, 
\begin{eqnarray}
&&{\cal L}_{\Delta_2} = \Delta^{++}_2
	 ( \bar \mu^c, \bar \tau^c ) \left ( \begin{array}{cc}
			 y_{22}\;\; & \;\;y_{23}\\
			\\
		y_{23}\;\; & \;\; y_{22}
		\end{array}
		\right ) L
		\left ( \begin{array}{c}
			 \mu \\ \tau
		\end{array} \right )
		\;,\quad y_{22}/y_{23}=10\;.
	\end{eqnarray}
In the following collider analysis, we employ the above couplings to perform the study with  omitting the  subscript “2”.

\section{Analysis of Four-Body Decay Results at Lepton Colliders}\label{section3}

\begin{figure}[]
    \centering
    \includegraphics[width=0.4\linewidth]{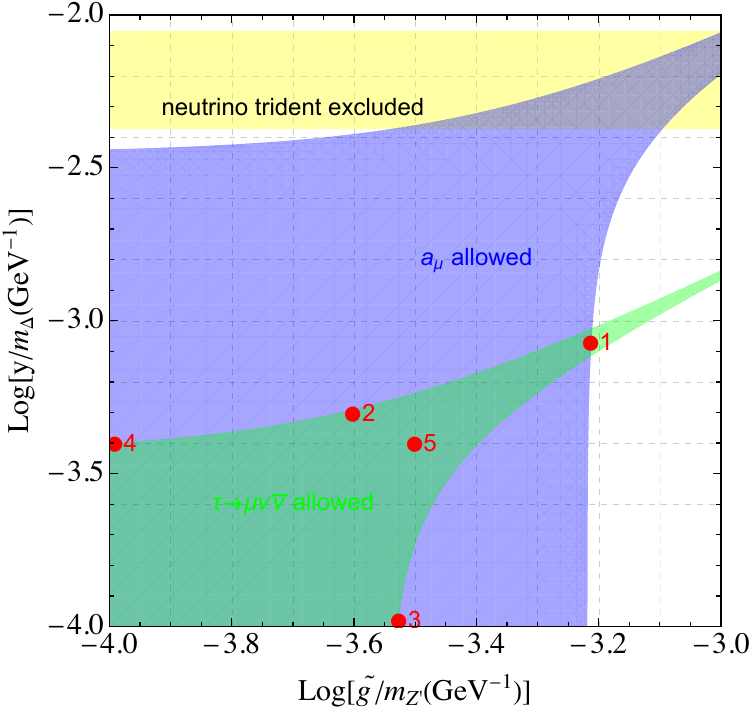}
    \caption{The allowed model parameter spaces in the  $\tilde g/m_{Z'}$-$y/m_{\Delta}$ plane. 
Different constraints are indicated using distinct colors: green for  $\tau \to \mu\nu\bar\nu$; yellow for neutrino trident process; and blue for muon $(g-2)$.  
Here the red markers denote the selected
benchmark points, chosen  to represent both typical and extreme parameter ratios of the $Z'$ boson and triplet scalars.
}
    \label{fig:range}
\end{figure}

Before presenting the detailed collider analysis, 
we first discuss the current existing bounds on the model parameters.
\begin{itemize}
\item $(g-2)_l$: One-loop exchange of $Z'$ and $\Delta^{+,++}$ both contribute to muon $(g-2)_\mu$,  with   opposite contributions as~\cite{Cheng:2021okr,Huang:2024iip}
\begin{eqnarray}\label{largeg-2}
\Delta a_{\mu}=\Delta a^{Z'}_{\mu}+ 
\Delta a^{\Delta}_{\mu}
=\frac{\tilde g_{Z'}^2 m_{\mu}^2}{12\pi^2m_{Z'}^2}\left(\frac{3m_{\tau}}{m_{\mu}}-2\right) -\frac{m_{\mu}^2}{16\pi^2}
   \frac{y_{22}^2}{m_{\Delta}^2}\;.
\end{eqnarray}
Here we adopt the limit $m_{\Delta,Z'}>>m_{\tau,\mu}$.
For $(g-2)_\mu$, FNAL reported a new experimental world average~\cite{Muong-2:2025xyk}, together with 
the updated theoretical prediction using lattice QCD calculations alone~\cite{Aliberti:2025beg},
yields a deviation of $a_\mu^{exp}-a_\mu^{SM}=39(64)\times 10^{-11}$.
Furthermore, the opposite contributions from $Z'$ and $\Delta$ cancel each other out, resulting in consistency with the recent muon $(g-2)_\mu$ measurement.
A $2\sigma$ confidence level is applied to constrain the model parameters in the $\epsilon - m_{Z'}$ plane, as indicated in blue in Fig.~\ref{fig:range}.
\item \textbf{$\tau \to \mu\nu\bar\nu$}: The decay processes are mediated by $Z'$ and $\Delta^+$ effects in terms of the ratios as~\cite{Cheng:2021okr,Huang:2024iip}
\begin{eqnarray}
     R^\tau&\equiv&\frac{\Gamma(\tau^-\to\mu^-\overline{\nu}_\mu\nu_\tau)}{\Gamma(\tau^-\to\mu^-\overline{\nu}_\mu\nu_\tau)_{\mathrm{SM}}}=\left(1-\frac{4m_W^2}{g^2}\frac{y_{22}^2}{4m_{\Delta^+}^2}\right)^2+\frac{4m_W^2}{g^2}\frac{\tilde g^2}{m_{Z^{\prime}}^2}\left(1-\frac{4m_W^2}{g^2}\frac{y_{22}^2}{4m_{\Delta^+}^2}
  \right)+ \frac{\tilde g^4}{g^4}\frac{4m_W^4}{m_{Z^{\prime}}^4}\;.
\end{eqnarray}
Combining the  SM predictions $R^\tau_{SM}= 0.972564\pm0.00001$~\cite{Pich:2009zza} and average experimental values $R^\tau_{aver}=0.972968\pm 0.002233$ from HFLAV collaboration~\cite{HFLAV:2022esi} and Belle-II~\cite{Corona:2024nnn}, the ratio can be obtained as  $R^\tau=1.00042\pm 0.00230$, which imposes the stringent bounds as shown in the  Fig.~\ref{fig:range}.
\item \textbf{$\nu_\mu+N\to \nu_{\mu,\tau} +N +\mu^+\mu^-$}:
The process known as  neutrino trident processes,  is induced by $\Delta^+$ at tree-level as~\cite{Cheng:2021okr,Huang:2024iip} 
\begin{eqnarray}\label{sigmaDelta^+}
   \frac{\sigma_{\Delta^+}}{  \sigma_{SM}}|_{trident}=
   \frac{\left(1+4 s^2_W- \frac{m^2_W}{g^2}\frac{2y_{22}^2}{m_{\Delta}^2}\right)^2+ 
    \left(1- \frac{m^2_W}{g^2}\frac{2y_{22}^2}{m_{\Delta}^2}\right)^2} 
   {[(1+4 s^2_W)^2 + 1]}\;,
\end{eqnarray}
Combining the current bounds from CHARM, CCFR and NuTeV~\cite{CHARM-II:1990dvf,CCFR:1991lpl,NuTeV:1999wlw}, we obtain the averaged value $0.95\pm0.25$. 
\end{itemize}

Motivated by the aforementioned constraints, we explore the viable parameter space of the $U(1)_{L_\mu - L_\tau}$ model as a potential explanation for the observed deviations in Fig.~\ref{fig:range}.
In order to better visualize the relevance between the lepton collider sensitivity and allowed regions, we define several benchmark points as

\begin{eqnarray}
 &&   B1: \tilde g/m_{Z'}(\text{GeV}^{-1})=0.62\times10^{-3}\;,\quad 
    y/m_{\Delta}(\text{GeV}^{-1})=0.83\times10^{-3}\;,\nonumber\\
&&     B2: \tilde g/m_{Z'}(\text{GeV}^{-1})=0.25\times10^{-3}\;,\quad y/m_{\Delta}(\text{GeV}^{-1})=0.5\times10^{-3}\;,\nonumber\\
&&     B3: \tilde g/m_{Z'}(\text{GeV}^{-1})=0.3\times10^{-3}\;,\quad \;
      y/m_{\Delta}(\text{GeV}^{-1})=0.1\times10^{-3}\;,\nonumber\\
&&     B4: \tilde g/m_{Z'}(\text{GeV}^{-1})=0.1\times10^{-3}\;,\quad \;
      y/m_{\Delta}(\text{GeV}^{-1})=0.4\times10^{-3}\;,
      \nonumber\\
&&     B5: \tilde g/m_{Z'}(\text{GeV}^{-1})=0.32\times10^{-3}\;,\quad 
      y/m_{\Delta}(\text{GeV}^{-1})=0.4\times10^{-3}\;.
\end{eqnarray}

These benchmarks exhibit varied combinations of the  gauge parameter $\tilde{g}/m_{Z'}$ and the triplet scalar parameter $y/m_\Delta$, ensuring both phenomenological viability and distinctive collider signatures.

\subsection{Four-Body final states at the Muon Colliders}

\begin{figure}[H]
    \centering 
    \begin{subfigure}{0.4\textwidth}
        \centering 
        \includegraphics[width=\linewidth]{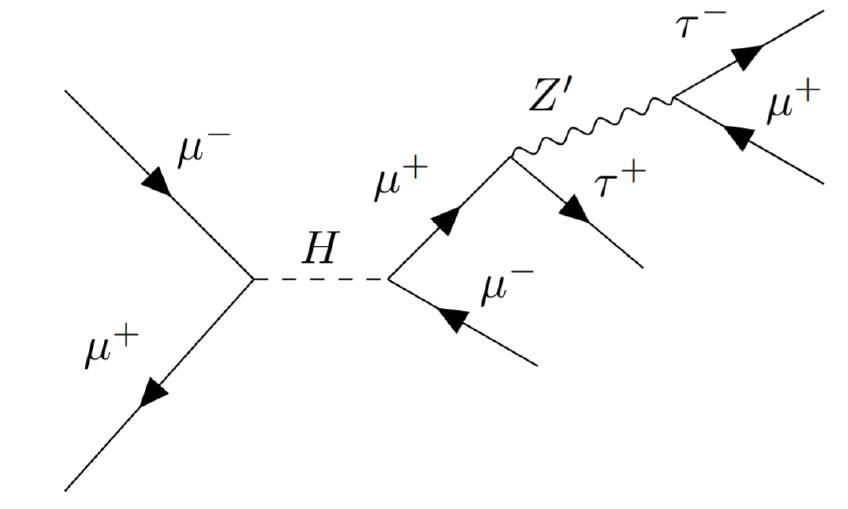}
    \end{subfigure}
    \quad
    \begin{subfigure}{0.4\textwidth}
        \centering 
        \includegraphics[width=\linewidth]{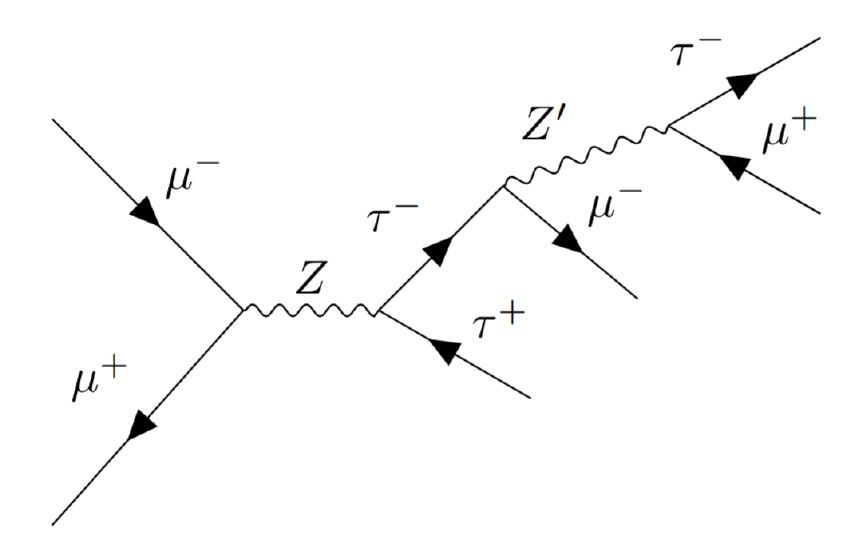}
    \end{subfigure}
    \caption{Representative Feynman diagrams for the four-lepton processes}
    \label{fig:feyn}
\end{figure}

In the maximally flavor-violating model $U(1)_{L_\mu-L_\tau}$, the process $\mu^-\mu^+ \rightarrow \mu^\pm \mu^\mp + \tau^\pm \tau^\mp$   can be mediated by the   gauge boson $Z'$ and new Higgs scalars, as shown in the typical Feynman diagram in Fig.~\ref{fig:feyn}.

\begin{figure}[H]
    \begin{subfigure}{0.48\textwidth}
    \includegraphics[width=\linewidth]{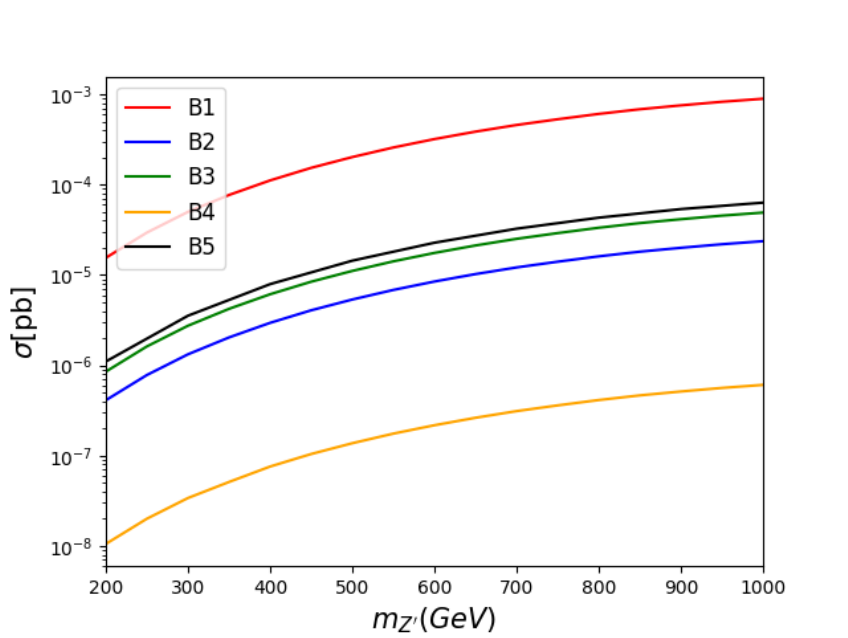}
    \end{subfigure}
    \begin{subfigure}{0.44\textwidth}
    \includegraphics[width=\linewidth]{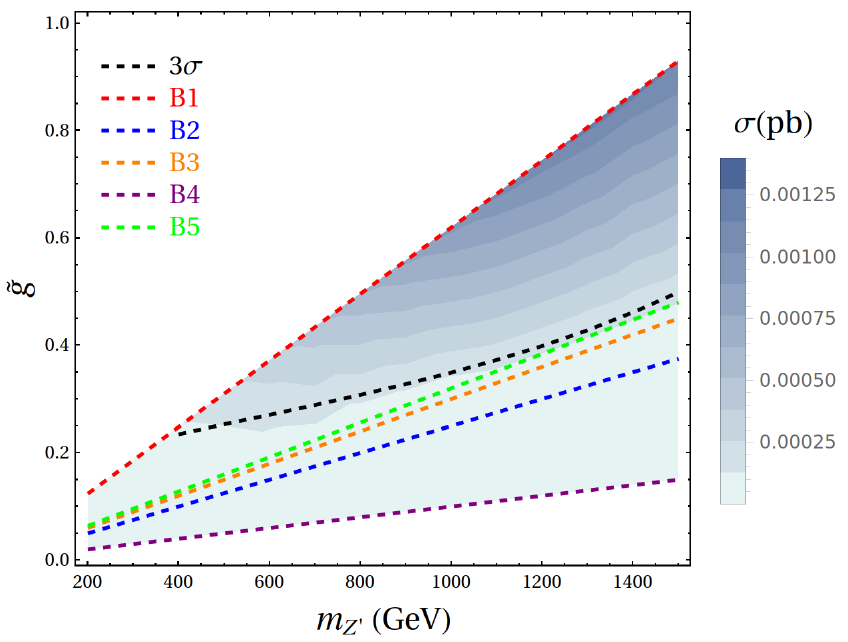}
    \end{subfigure}
    \caption{ (a) The  cross section $\sigma(\mu^-\mu^+ \rightarrow \mu^\pm \mu^\mp + \tau^\pm \tau^\mp)$ as a function of $m_{Z'}$ at a muon collider with $\sqrt{s}=3$ TeV. 
    (b) The  signal cross-section in the plane $\tilde g-m_{Z'}$.}
    \label{fig:2}
\end{figure}

For the benchmark points mentioned above,  we calculate the production cross sections $\sigma(\mu^-\mu^+ \rightarrow \mu^\pm \mu^\mp + \tau^\pm \tau^\mp)$ as a function of $m_{Z'}$ at a muon collider, as presented in the left panel of Fig.~\ref{fig:2}. 
We find that the corresponding cross sections increase monotonically with $m_{Z'}$, reflecting the enhanced phase space and coupling strengths at higher resonance masses.
 Benchmark B1 yields the largest cross sections, reaching $10^{-3}$ pb for $m_{Z'} \sim 1$ TeV, while B4 predicts significantly suppressed cross sections, with a suppression of approximately three orders of magnitude over the same mass range.
These differences illustrate the strong dependence of collider phenomenology on the underlying parameter choices, highlighting the importance of multi-lepton final states as sensitive probes of  this model.

For a collider center-of-mass energy of 3 TeV, we  assume an integrated luminosity of $1~\mathrm{ab}^{-1}$ \cite{Aime:2022flm}.
Since this process conserves both charged lepton number and lepton flavor, it receives  SM background contributions, with the value $8.19\times10^{-4}$ pb.
In order to extract signatures more effectively, we can   adopt the following criteria as~\cite{Cowan:2010js} 
\begin{equation}
\frac{S}{\sqrt{B+S}}=\frac{\sigma\times\mathscr{L}}{\sqrt{(\sigma_{SM}+\sigma)\times\mathscr{L}}}.
\label{eq:6}
\end{equation}
The corresponding values are summarized in Table.~\ref{tab:1},  which presents benchmark scenarios B3, B2, and B1 with different combinations of the effective gauge parameters $\tilde{g}/m_{Z'}$ and triplet parameters.
 For each scenario, we report the predicted signal cross section $\sigma$ and the corresponding statistical significance.

\begin{table}[H]
 \caption{The cross section and statistical significance for different benchmarks at the  luminosity of  $\mathcal{L}=1\text{ab}^{-1}$. }
    \centering
    \begin{tabular}{c|c|c|c|c|c|c|c}
    \hline
    \hline
    &$\tilde{g}/m_{Z^\prime}(10^{-3}\text{GeV}^{-1})$ &$m_{Z^\prime}(\text{GeV})$&$\tilde{g}$ & $m_{\Delta}(\text{GeV})$ & $Y_{22}$ &$\sigma\times10^{-4}$(pb)  &S/$\sqrt{B+S}$  \\
    \hline
    \hline   
 \multirow{2}{*}{B2} &\multirow{2}{*}{0.25} & 500 & 0.125 & \multirow{2}{*}{500} & \multirow{2}{*}{0.25} & 0.05 & 0.19 \\
\cline{3-4} \cline{7-8} 
                      & & 800 & 0.20&    &                  & 0.16 & 0.56 \\ 

  \hline   
  \multirow{2}{*}{B3}&\multirow{2}{*}{0.3} & 500 & 0.15 &  \multirow{2}{*}{500}&\multirow{2}{*}{0.05} & 0.11&0.39  \\
 \cline{3-4} \cline{7-8}  
 & & 800 & 0.24 & &  &0.33   &  1.15 \\

 \hline   
  \multirow{2}{*}{B1}&\multirow{2}{*}{0.62} & 500 & 0.31 &  \multirow{2}{*}{500}&\multirow{2}{*}{0.415} & 2.03& 6.35 \\
   \cline{3-4} \cline{7-8}   
 & & 800 & 0.496& &  &6.10  &  16.14 \\

    \hline
    \hline
    \end{tabular}
    \label{tab:1}
\end{table}

The corresponding cross section can be projected into the plane of $\tilde g-m_{Z'}$ as shown in the right panel of Fig.~\ref{fig:2}. 
 The shaded contours denotes the cross section  as a function of the  mass $m_{Z'}$ and the  coupling $\tilde{g}$, with darker regions corresponding to larger signals.
Benchmark points B1-B5 are indicated by distinct colored dashed lines, while the black dashed line marks the $3\sigma$ sensitivity threshold.
Among the benchmarks, B1 lies mostly above the
 $3\sigma$ curve for $m_{Z'}>400$ GeV, indicating strong discovery prospects. 
 In contrast, B2-B5 remain well below the observable threshold. 
 Therefore, the $Z'$ signal could be observable in the high-mass regime at a muon collider under scenario B1.
 Additionally, we further check that the contribution from the triplet scalar has only a minor effect on the overall cross section, suggesting that the gauge sector predominantly determines the collider sensitivity in this channel.


The above analysis is based on the unpolarized case. In practice,  the initial-state lepton beams at a muon collider can be polarized in various configurations, as discussed in Refs.~\cite{CMB-HD:2022bsz,Chen:2017yel}. In this study, we consider two representative polarization scenarios:
  $\{P_{\mu^+}, P_{\mu^-}\} = \{-20\%, 80\%\}$ and $\{80\%, -80\%\}$. 
Under these configurations, the signal significance exhibits notable improvements, as summarized in Table. \ref{tab:2}.
For $\{P_{\mu^+}, P_{\mu^-}\} = \{-20\%, 80\%\}$, 
the significance increases by approximately 14\% for benchmark B3 and 11\% for B1, relative to the unpolarized case.
For $\{P_{\mu^+}, P_{\mu^-}\}=\{80\%, -80\%\}$, 
the improvement is more substantial—about 64\% for B3 and 50\% for B1.
These enhancements arise because suitable initial-state polarization can suppress background processes that preferentially couple to a specific chirality, while simultaneously enhancing the signal cross section.
In particular, a large asymmetry between $P_{\mu^+}$ and $P_{\mu^-}$ maximizes the interference between the SM and $Z'$ amplitudes, thereby boosting the sensitivity in both the cross section and angular observables. 

\begin{table}[htbp]
   \caption{Same as Table.~\ref{tab:1}, but for different beam polarization configurations.}
    \centering
    \begin{tabular}{c|c|c|c|c|c|c|c|c|c|c|c}
        \hline
        \hline
         & \multirow{2}{*}{$\tilde{g}/m_{Z^{\prime}}(10^{-3}\text{GeV}^{-1})$} &\multirow{2}{*}{$m_{Z^{\prime}}$} & \multirow{2}{*}{$\tilde{g}$}& \multirow{2}{*}{$m_{\Delta}(\text{GeV})$}& \multirow{2}{*}{$Y_{22}$} & \multicolumn{2}{c|}{\{$P_{\mu^+}$,$P_{\mu^-}$\}=\{0\%,0\%\}} & \multicolumn{2}{c|}{\{$P_{\mu^+}$,$P_{\mu^-}$\}=\{-20\%,80\%\}} & \multicolumn{2}{c}{\{$P_{\mu^+}$,$P_{\mu^-}$\}=\{80\%,-80\%\}} \\
        \cline{7-12}
       & & & & & &$\sigma\times10^{-4}(pb)$ & $S/\sqrt{B+S}$ & $\sigma\times10^{-4}(pb)$ & $S/\sqrt{B+S}$ & $\sigma\times10^{-4}(pb)$ & $S/\sqrt{B+S}$ \\
        \hline
        
          \multirow{3}{*}{B3}&\multirow{3}{*}{0.3} & 600 & \multirow{1}{*}{0.18} & \multirow{3}{*}{500} & \multirow{3}{*}{0.05}&0.18 &0.61 & 0.20&0.7 &0.29 &1.00  \\
       
        & & 800 & \multirow{1}{*}{0.24} & &  &0.33 &1.14 &0.38 & 1.31& 0.57& 1.88 \\
        
         & & 1000 & \multirow{1}{*}{0.3}& &  &0.49 &1.67 & 0.56&1.91&0.82&2.73 \\

         \hline
        
       \multirow{3}{*}{B1}& \multirow{3}{*}{0.62}&600 & \multirow{1}{*}{0.36}  & \multirow{3}{*}{500} & \multirow{3}{*}{0.415}  &2.83 &8.53 &3.23 &9.55&4.68&13.05 \\
        
        & &800 & \multirow{1}{*}{0.48} & & & 5.35
         & 14.53&6.13&16.20 &8.90&21.53  \\
         
         & &1000 & \multirow{1}{*}{0.6} &  & &7.89 &19.67 &9.01&21.72&14.07&29.82\\
        \hline
        \hline
    \end{tabular}
    \label{tab:2}
\end{table}

To systematically investigate the implications of our model, we analyze the forward-backward asymmetry $A_{\text{FB}}$ defined by~\cite{ALEPH:2005ab} 
\begin{equation}
A_{FB}=\frac{\sigma_F-\sigma_B}{\sigma_F+\sigma_B},\quad 
\sigma_F=\int^1_0\frac{d\sigma}{dcos\theta}dcos\theta, \quad \sigma_B=\int^0_{-1}\frac{d\sigma}{dcos\theta}dcos\theta .
\label{eq:7}
\end{equation}
Here the scattering angle $\theta$ is defined as the angle between the outgoing $\mu$ and $\tau$ in the center-of-mass frame of the collision.


\begin{figure}[H]
\centering
    \includegraphics[width=0.45\linewidth]{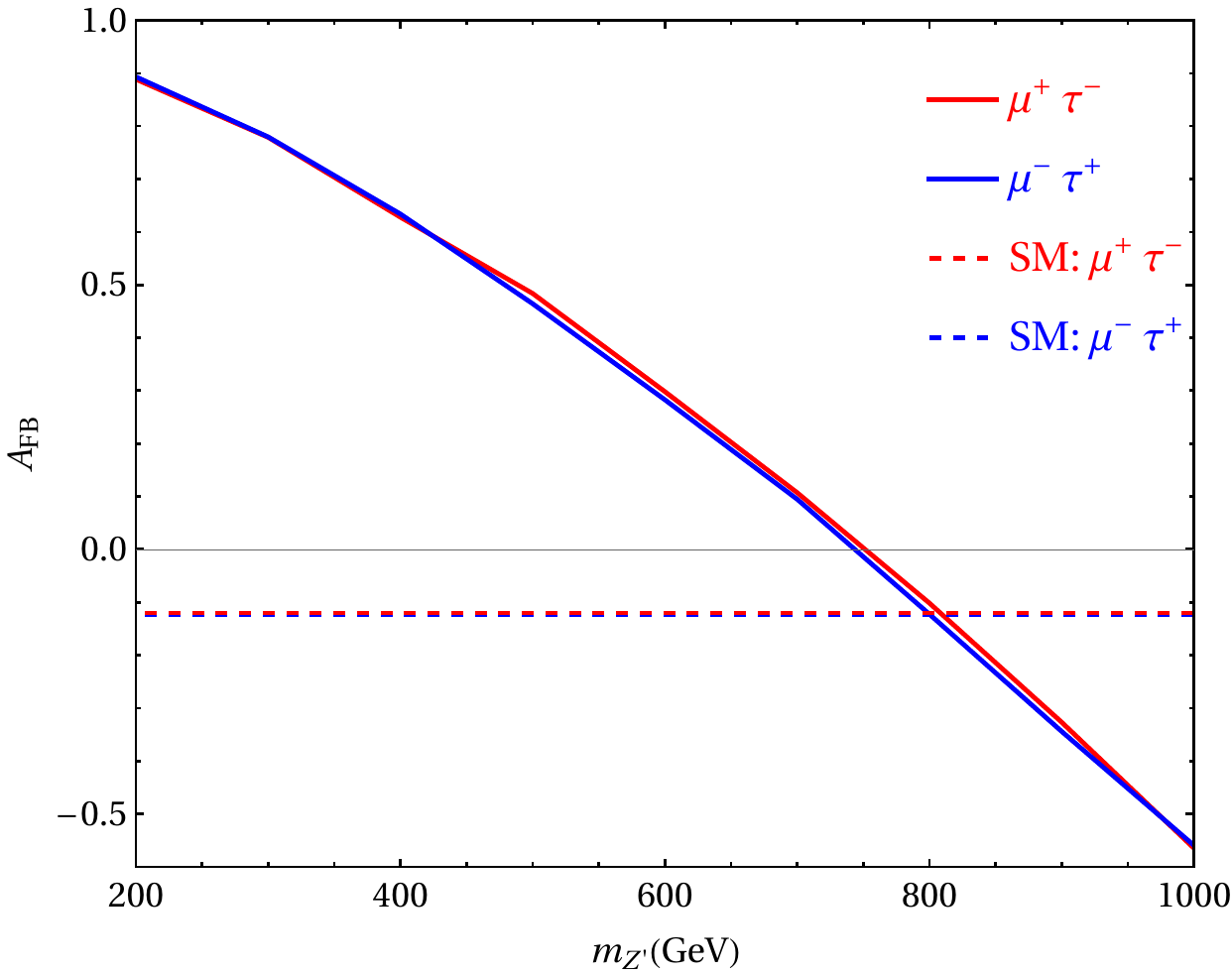}
    \caption{ The forward-backward asymmetry $A_{FB}$ as the function with  $m_{Z^\prime}$ for $\tilde{g}=0.1$ with $m_{\Delta}=500 $ GeV and $Y_{22}=0.415$. 
      }
       \label{fig:8}
\end{figure}

Firstly, we focus on the evolution of the forward-backward asymmetry $A_{\text{FB}}$ with respect to the mass $m_{Z'}$, considering the promising benchmark point B1 discussed earlier. 
By fixing the triplet scalar mass $m_{\Delta}=500 $ GeV, we can isolate the pure impact of the 
$Z^\prime$ mass on the asymmetry.  Within the SM, the process $\mu^{+}\mu^{-} \to \mu^{\pm}\mu^{\mp}\tau^{\pm}\tau^{\mp}$ is governed by $\gamma$/Z exchange and yields $A_{FB}^{SM}\sim 0$, due to the absence of chiral flavor violating interference.
 For $\tilde{g} = 0.1$, the allowed region in Fig.~$\ref{fig:range}$ corresponds to a $m_{Z^\prime}$ range of 200-1000 GeV,  As $m_{Z^\prime}$ increases, the predicted forward-backward asymmetry $A_{FB}$ decreases monotonically and approaches zero around $m_{Z^\prime} \simeq 750$ GeV, as illustrated in Fig.~$\ref{fig:8}$
This monotonic decrease reflects the suppression of parity-violating interference effects as the 
$Z^\prime$ becomes heavier. 
Notably, the transition from positive to negative asymmetry is a characteristic feature of the model and could serve as a discriminating observable in the search for heavy 
$Z'$ bosons at future colliders.
Furthermore, the two curves corresponding to the final states $\mu^- \tau^+$ (blue) and  $\mu^+ \tau^-$ (red) nearly overlap, corroborating the fact that the forward-backward asymmetry is largely insensitive to the electric charge assignment of the final-state leptons.
Note that the behavior of the forward-backward asymmetry in our analysis is a direct consequence of the maximal flavor-violating $Z'$ structure.
The maximally flavor-violating scenario in our model arises from the discrete exchange symmetry, $Z' \to -Z'$ and $1 \leftrightarrow 2$, which is guaranteed even at one-loop level due to the  new charge assignment.
 This exchange symmetry may be broken by quantum corrections at higher loop levels, leading to small flavor-diagonal couplings to muons and taus. However, the resulting flavor-diagonal couplings $\tilde g_{\mu\mu}$ are much smaller than the flavor-changing couplings $\tilde g_{\mu\tau}$, so that the dominant contribution to $A_{FB}$ comes from the flavor-changing couplings, and the corresponding results remain largely unaffected.

\begin{figure}[H]
    \centering
    \captionsetup[subfigure]{labelformat=simple, labelsep=space}
    \renewcommand{\thesubfigure}{FIG.\thefigure(\alph{subfigure})} 

    \begin{subfigure}{0.48\textwidth}
        \includegraphics[width=\linewidth]{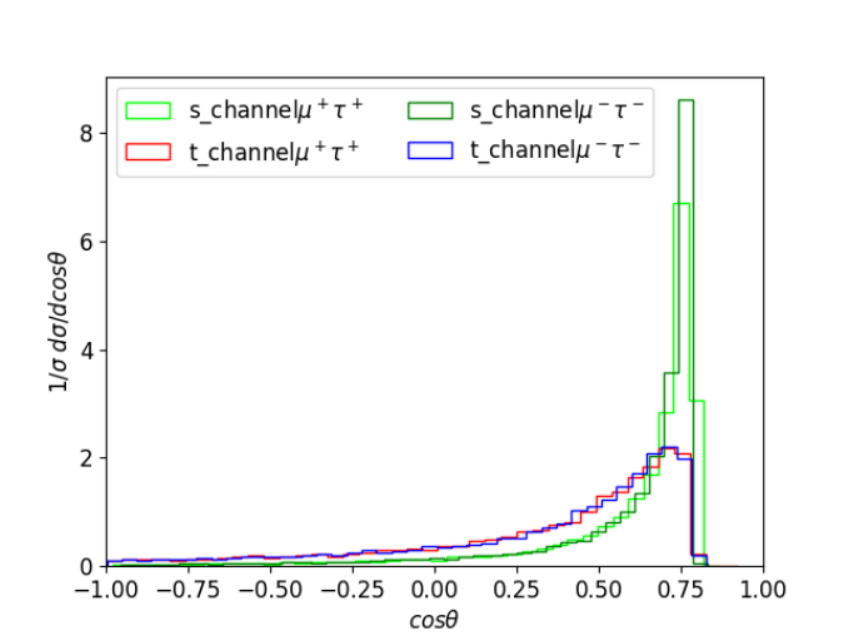}
        
        \label{fig:6a} 
    \end{subfigure}
    \hspace{0.05cm}
    \begin{subfigure}{0.48\textwidth}
        \includegraphics[width=\linewidth]{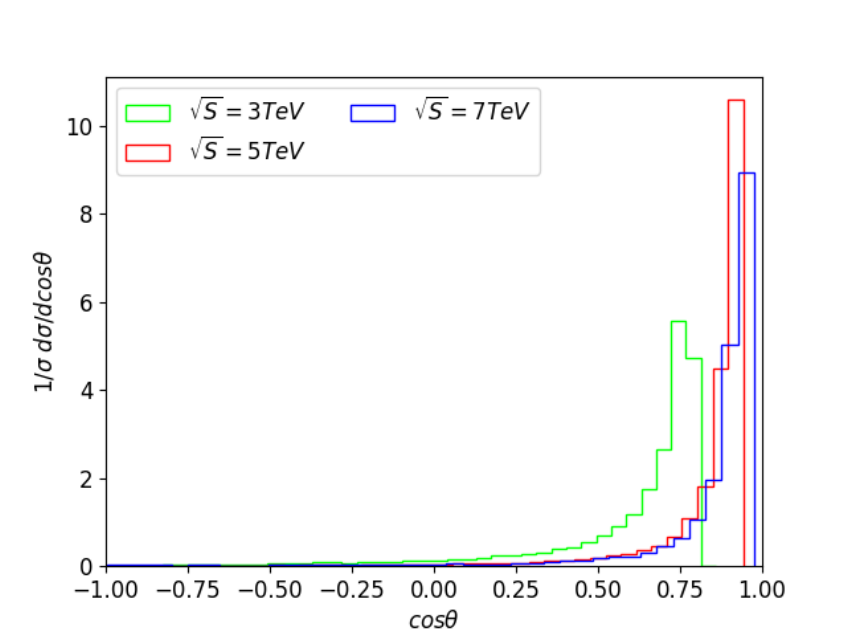}
 
        \label{fig:6b} 
    \end{subfigure}
    \caption{(a) The angular distribution at a center-of-mass energy of 3 TeV for different channels. (b) The angular distribution with different collider  energies.}
    \label{fig:6}
\end{figure}

Furthermore, we aim to investigate the individual effects of doubly charged scalars.
Since the $Z'$ boson exhibits only vector-current interactions in our model, we can probe the chiral properties of the doubly charged scalars by defining 
\begin{eqnarray}
 \mathcal{R}=\frac{\sigma_{LR}+\sigma_{LL}-\sigma_{RR}-\sigma_{RL}}{\sigma_{LR}+\sigma_{LL}+\sigma_{RR}+\sigma_{RL}}\;.
\end{eqnarray}
Here $\sigma_{ij}$, where $i,j=L,R$,  denote 
the polarized cross sections, with “L” and “R” denoting 100\% left- and right-handed polarization, respectively.
A numerical evaluation yields
 $\mathcal{R} \simeq -1$,
indicating a maximal left-right asymmetry ($100\%$).
This also serves as a cross-check of the interaction structure, where the left-handed vector current (V-A) is mediated by the triplet scalar.
To further investigate the details, we similarly analyze the angular distribution of the differential cross section, as shown in Fig.~\ref{fig:6}.
Here $\theta$ is defined as the angle between the two final-state charged leptons originating from the decay of the doubly charged Higgs boson.
In the left panel, four different final-state configurations are presented, revealing that the angular distribution is truncated around $\cos\theta = 0.8$. This is because the difficulty in distinguishing the final-state $\mu$ and $\tau$ when the opening angle is small.
The structure indicates that $A_{FB}$ tends to be positive, with corresponding values of 0.8772, 0.6216, 0.8736, and 0.622, respectively.
In the right panel, the angular distribution is shown for different collider energies, 
revealing that lower energies correspond to a larger truncation angle in the angular distribution.
Additionally, when the center-of-mass energy reaches 6 TeV, the angular distribution becomes visible across the full range of $\cos \theta\in [-1,1]$.
Although the initial design stage plans to operate at a maximum energy of 3 TeV,
a 10 TeV scenario is foreseeable in the future, provided that challenges such as muon beam cooling and the particle’s short lifetime are resolved.

\begin{figure}[H]
\centering
    \includegraphics[width=0.45\linewidth]{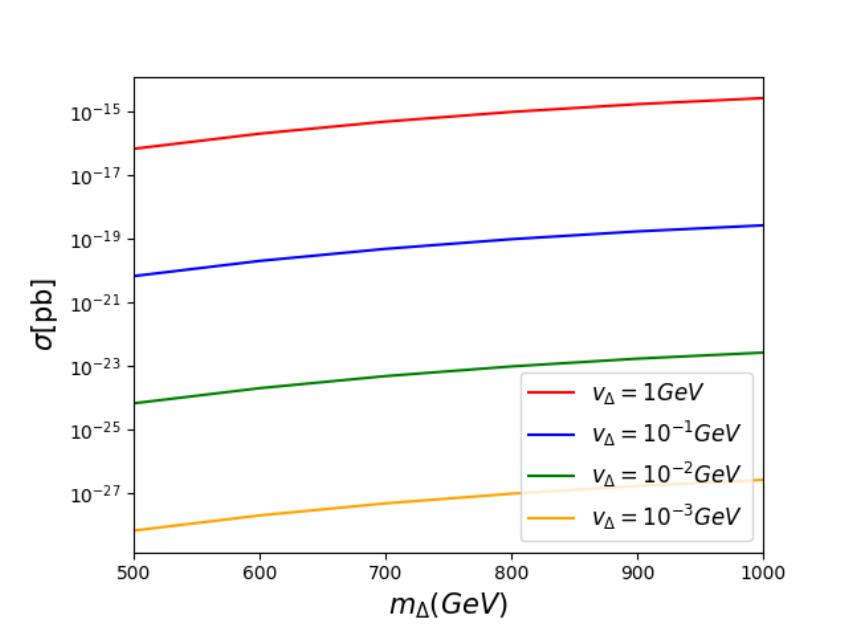}
    \caption{  The cross section for the process $\mu^+\mu^-\to \Delta^{++}\Delta^{--}\to 4W$ as a function of the mass  $m_\Delta$ under different values $v_{\Delta}$. 
      }  
       \label{fig:0712}
\end{figure}

In addition to the charged lepton final states, 
the doubly-charged scalar can decay into a pair of $W^\pm$ bosons, with the corresponding expressions as~\cite{Huang:2025osf}
\begin{eqnarray}
\frac{\Gamma(\Delta^{++}\to W^+W^+)}{\Gamma(\Delta^{++}\to \ell^+\ell^+)}
&\approx& \frac{v_{\Delta}^2}{v_{SM}^4}\frac{ m_{\Delta}^2 }{ y^2} 
\approx 10^{-10}\left(\frac{v_{\Delta}}{1\ \text{GeV}}\right)^2
\left(\frac{250\ \text{GeV}}{v_{SM}}\right)^4
\left(\frac{m_\Delta/\text{GeV}}{y}\right)^2 ,
\label{vev}
\end{eqnarray}
Here $v_{\text{SM}}=246$ GeV.
We find that the decay into a pair of $W^\pm$ 
primarily depends on the triplet vacuum expectation value $v_\Delta$, while  the decay into charged leptons is predominantly governed by the Yukawa coupling.
For the $4W$ boson final states originating from $\Delta^{\pm\pm}$ decays, the production  cross section with $m_{\Delta}$ is presented in Fig.~\ref{fig:0712},  which explicitly demonstrates the significant role of $v_\Delta$.
Furthermore, the cross section exhibits a monotonic increase as $m_{\Delta}$ increases.
Additionally, the cross section decreases by several orders of magnitude when decreasing $v_{\Delta}$ from 1 to $ 10^{-3} $ (corresponding to the red and orange curves, respectively).
This behavior is crucial for predicting detectability in collider experiments:
higher values of $v_{\Delta}$ and $m_{\Delta}$
 increase the cross section to levels potentially observable above background noise.

    
    

\subsection{Four-Body final states at electron-positron Colliders}

The muon collider analysis in the previous subsection can also be applied to the case of an electron collider.
To facilitate comparison, we choose the same center-of-mass energy $\sqrt{s} = 3$ TeV for the electron-positron collider, as reported in the CILC design~\cite{Aicheler:2018arh}.
The corresponding  integrated luminosity is $3000~\text{fb}^{-1}$.

\begin{figure}[H]
    \begin{subfigure}{0.48\textwidth}
    \includegraphics[width=\linewidth]{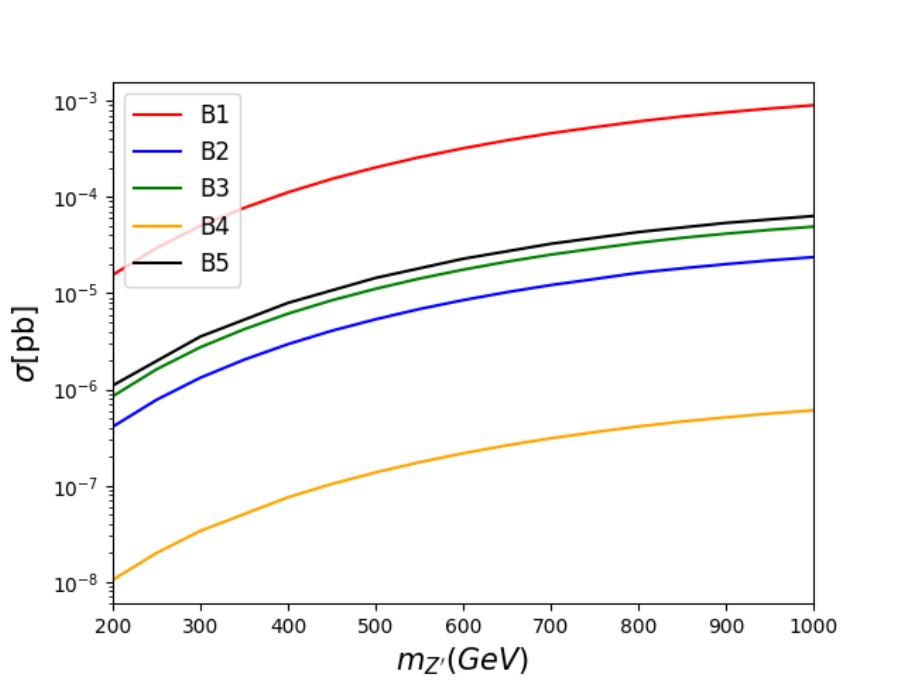}
    \end{subfigure}
    \begin{subfigure}{0.44\textwidth}
    \includegraphics[width=\linewidth]{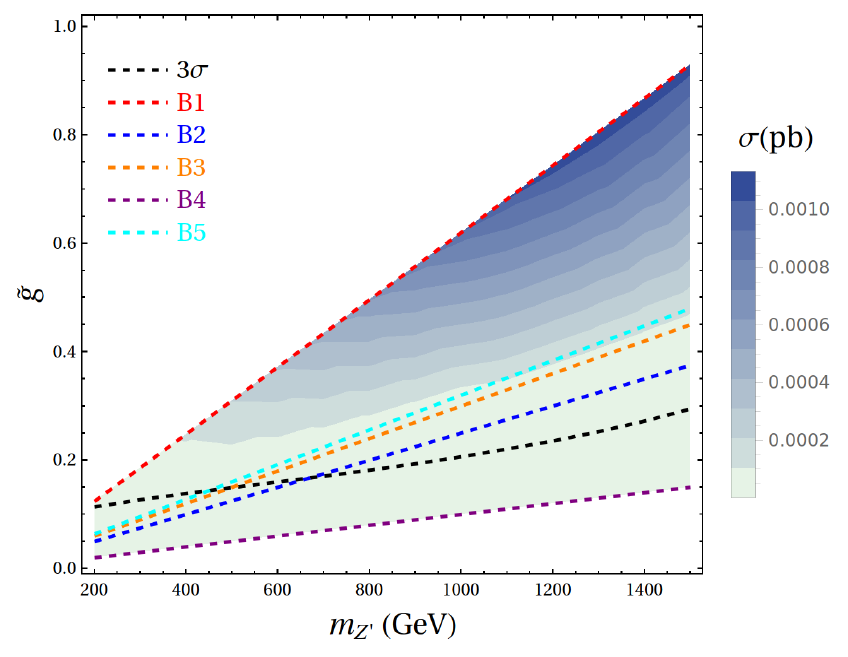}
    \end{subfigure}
    \caption{  Same as Fig.~\ref{fig:2} but for $\sigma(e^-e^+ \rightarrow \mu^\pm \mu^\mp + \tau^\pm \tau^\mp)$.
    }
    \label{fig:10}
\end{figure}

Note that since our model framework involving $Z'$ and $\Delta^{\pm\pm}$ only couples to muons and taus, the electron collider contributes only through $s$-channel processes to produce $\mu^+ \mu^- \tau^+ \tau^-$.
The corresponding cross sections for four different benchmark points are indicated in the left panel of Fig.~\ref{fig:10}.
A similar increasing trend is observed, consistent with the behavior in Fig.~\ref{fig:2}.
Additionally, benchmark B1 yields the largest cross section, approximately $\mathcal{O}(10^{-3})$ pb, exceeding those of the other three benchmarks.


For the process $e^+e^- \to \mu^{\pm}\mu^{\mp}\tau^{\pm}\tau^{\mp}$,
the SM background  arises from purely $\gamma/Z$-mediated processes,  $2.92\times10^{-5}$ pb, while the signal arises from diagrams involving $Z^\prime$ or doubly charged scalar exchange.
Adopting the statistical significance defined in Eq.~\ref{eq:6}, we can obtain the projected cross section  on the plane of   $m_{Z^\prime}-\tilde{g}$, as shown in  the right panel of Fig.~\ref{fig:10}.
We find that the cross section increases with both $m_{Z'}$ and $\tilde{g}$,
reaching a maximum at their largest values.
Similarly, four benchmark scenarios (B1-B5) are included for comparison, along with the corresponding $3\sigma$ discovery contour.
Among these benchmark scenarios, B1 lies well above the $3\sigma$ discovery threshold across the entire mass range, indicating the highest potential for an observable signal, consistent with the results from the muon collider case.
The only difference compared to the muon collider case is that B2, B3 and B5 can exceed the $3\sigma$ discovery threshold at large values of $m_{Z'}$.
Specifically, B2 exceeds the threshold when $m_{Z'} \gtrsim 650$ GeV,  B3 does so for $m_{Z'} \gtrsim 500$ GeV, and B5 does so for $m_{Z'} \gtrsim 450$ GeV.
These results underscore the strong dependence of new physics discovery prospects on the choice of parameter space, particularly in the context of electron collider scenarios.

\begin{figure}[H]
\centering
    \includegraphics[width=0.45\linewidth]{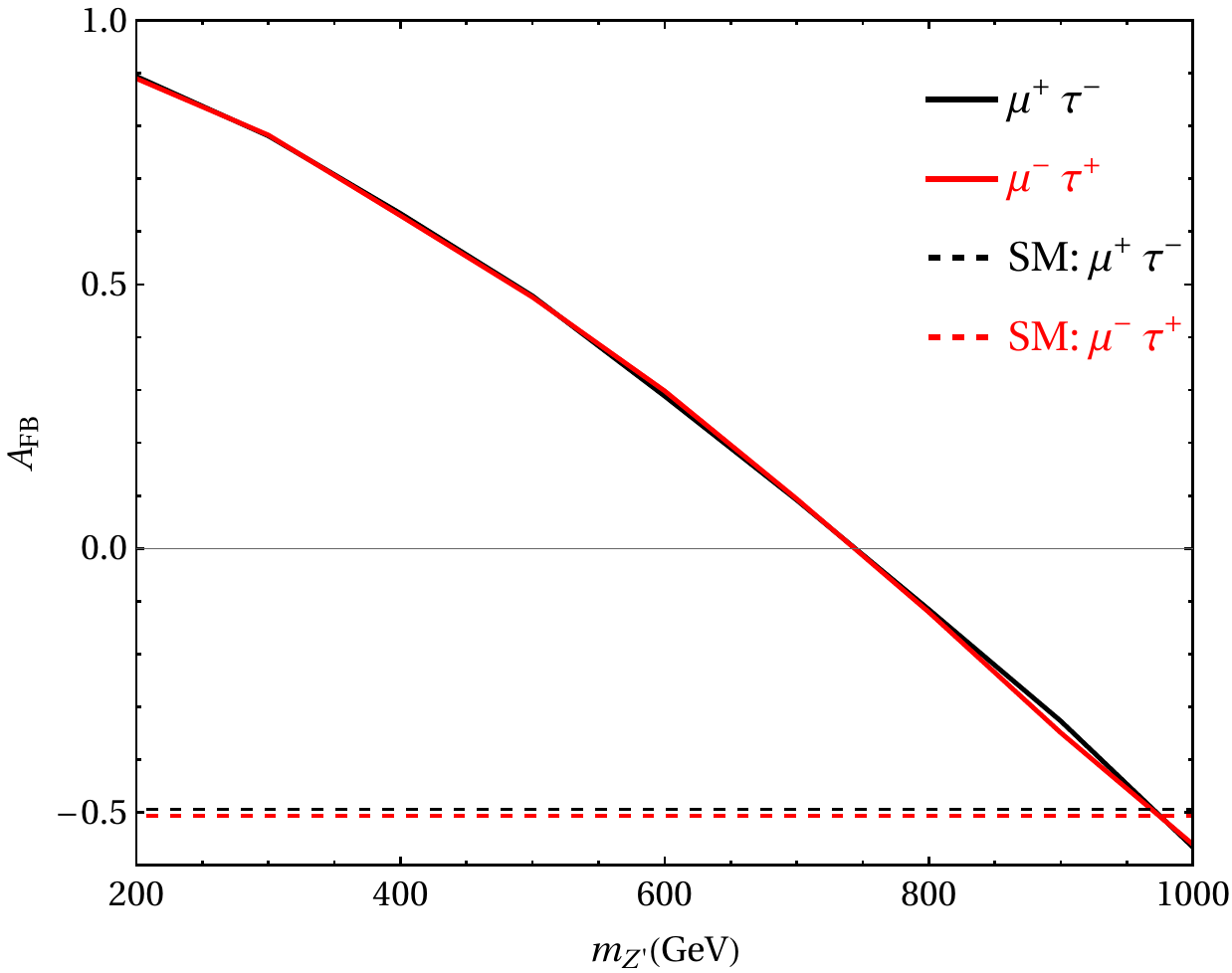}
    \caption{ The forward-backward asymmetry $A_{FB}$ as the function with  $m_{Z^\prime}$ for $\tilde{g}=0.1$ with $m_{\Delta}=500 $ GeV and $Y_{22}=0.415$. 
      } 
       \label{fig:neweafb}
\end{figure}

Similarly, the forward-backward asymmetry is investigated through the analysis of the angular distribution, as shown in Fig.~\ref{fig:neweafb}.
A clear trend is observed with increasing mass $m_{Z^\prime}$, which is of great significance for understanding the underlying physical mechanisms and guiding experimental observations.



\section{Conclusion}
\label{section4}

In this work, we investigated the collider phenomenology of the $U(1)_{L_\mu-L_\tau}$ model with maximal flavor violation, concentrating on four-lepton final states as a sensitive probe of new physics. We demonstrated that 
the signal cross sections increase with both the effective parameters $\tilde{g}/m_{Z^\prime}$ and the mass $m_{Z^\prime}$, 
whereas the contribution from triplet Yukawa couplings remains comparatively small within the phenomenological allowed regions.
We further demonstrated that the forward-backward asymmetry constitutes a robust observable, with its monotonic dependence on $m_{Z^\prime}$ serving as a distinctive signature of the model.
Polarization of the initial beams enhances sensitivity by suppressing Standard Model backgrounds and amplifying chiral effects associated with the new interactions, thereby facilitating the study of triplet scalar effects.
Among the benchmark scenarios considered, the B1 point yields the highest discovery significance, exceeding the $3\sigma$
 threshold at both muon and electron-positron colliders.
 The B2 and B3 scenarios also offer good sensitivity at electron colliders for large values of $m_{Z'}$.
 The difference between electron and muon colliders lies in the presence of additional t- and u-channel contributions in the muon collider, in contrast to the electron collider, which only involves s-channel processes.

Overall, our study demonstrates that high-energy lepton colliders can offer clear and testable signatures of the $U(1)_{L_\mu-L_\tau}$ framework.
Multi-lepton final states, in conjunction with polarization and angular observables, provide experimentally accessible handles to probe lepton flavor violation and to explore the underlying gauge and scalar dynamics of the model.

\section*{Acknowledgments}
 The work is supported by National Natural Science Foundation of China under Grant No.12505112 and 11805081, Shandong Province Natural Science Foundation under Grant No.ZR2024QA138, ZR2022MA056 and State Key Laboratory of Dark Matter Physics, University of Jinan Disciplinary Cross-Convergence Construction Project 2024 (XKJC- 202404)
and Jin Sun is supported by IBS under the project code, IBS-R018-D1.


\end{document}